\documentstyle[multicol,aps]{revtex}
\renewcommand{\narrowtext}{\begin{multicols}{2}
\global\columnwidth20.5pc\noindent}
\renewcommand{\widetext}{\end{multicols}
\global\columnwidth42.5pc}
\begin{document}
\draft
\preprint{22 September 1998}
\title{Magnetic Properties of Quantum Ferrimagnetic Spin Chains:\\
       Ferrimagnetism $\approx$ 
       Ferromagnetism $+$ Antiferromagnetism$\,$?}
\author{Shoji Yamamoto}
\address{Department of Physics, Okayama University,
         Tsushima, Okayama 700-8530, Japan}
%\date{Received \hspace{6cm}}
\date{Received 22 September 1998}
\maketitle
\begin{abstract}
   Magnetic susceptibilities of spin-$(S,s)$ ferrimagnetic
Heisenberg chains are numerically investigated.
It is argued how the ferromagnetic and antiferromagnetic features
of quantum ferrimagnets are exhibited as functions of $(S,s)$.
Spin-$(S,s)$ ferrimagnetic chains behave like combinations of
spin-$(S-s)$ ferromagnetic and spin-$(2s)$ antiferromagnetic chains
provided $S=2s$.
\end{abstract}
\narrowtext

   Quantum behavior of mixed-spin systems is one of the hot topics and
a recent progress in the theoretical understanding of it deserves
special mention.
In contrast with pioneering calculations \cite{Dril13} in the 1980s,
the current vigorous argument might more or less be motivated by the
Haldane conjecture \cite{Hald64} in that it covers not only
ferrimagnetism but also mixed-spin antiferromagnetism.
As coexistent spins of different kinds do not necessarily result in
magnetic ground states, it remains a stimulative problem there
whether the system is massive or massless.
Actually various mixed-spin chains \cite{Fuku09,Taka} and ladders
\cite{Fuku98} with singlet ground states have also been discussed
extensively via the nonlinear-$\sigma$-model technique with particular
emphasis on the competition between massive and massless phases.
On the other hand, an arbitrary alignment of alternating spins $S$ and
$s$, which is described by the Hamiltonian
\begin{equation}
   {\cal H}
      =\sum_j
       \left(
        J_1\mbox{\boldmath$S$}_{j} \cdot \mbox{\boldmath$s$}_{j}
       +J_2\mbox{\boldmath$s$}_{j} \cdot \mbox{\boldmath$S$}_{j+1}
       \right)\,,
   \label{E:H}
\end{equation}
shows ferrimagnetism instead of antiferromagnetism and
thus has gapless excitations from the ground state.

   Chemical explorations into ferrimagnetic chain compounds have
vigorously been reported so far.
In an attempt to obtain bimetallic complexes including one-dimensional
systems, CuMn(S$_2$C$_2$O$_2$)$_2$(H$_2$O)$_3$$\cdot$4.5H$_2$O
\cite{Glei73} was first synthesized.
Since then numerous bimetallic chain compounds \cite{Pei38}
were systematically synthesized and measured.
In the series of investigations, the smaller spin $s$ was fixed to
$1/2$ and the larger spin $S$ was changed from $1/2$ to $5/2$
featuring several important concepts.
One of their main interests was the achievement of a ferromagnetic
interaction between nearest-neighbor metal ions, namely, the design
of molecular-based ferromagnets.
Therefore quantum ferrimagnetism was not necessarily the main subject
there.
Though there also appeared theoretical investigations
\cite{Dril13,Dril53} in close contact with experiments, the argument
was restricted to the spin combinations $(S,1/2)$ and the larger spin
$S$ was sometimes treated as classical.

   In such circumstances, several renewed interests
\cite
{Alca67,Pati94,Kole36,Breh21,Nigg31,Yama09,Ono76,Kura29,Yama08,Ivan14,Mais},
have recently been introduced into the field.
Alcaraz and Malvezzi \cite{Alca67} predicted the universal low-energy
physics of anisotropic ferrimagnetic chains and the distinct behavior
of isotropic ones.
Kuramoto \cite{Kura29} discussed critical properties of the model in a
field showing its magnetization curve with a plateau.
Maisinger {\it et al.} \cite{Mais} revealed the field-induced
double-peak structure of the specific heat.
Another noteworthy subject, which is the main interest here, is the
coexistence of the ferromagnetic and antiferromagnetic features in
quantum ferrimagnets.
The author {\it et al.} pointed out in their recent study
\cite{Breh21,Yama09,Yama08} on the model (\ref{E:H}) with
$(S,s)=(1,1/2)$ that at low temperatures the model acts as a
ferromagnet, while at intermediate temperatures it behaves like a
gapped antiferromagnet.
In this article, we quantitatively verify to what extent this scenario
is valid discussing various combinations of $(S,s)$.
Even though there is an ambiguous belief that the $(S,s)=(1,1/2)$ case
is representative of all the Heisenberg ferrimagnetic chains, it is
not yet really clear whether or how the quantum behavior of the model
depends on the constituent spins.
We here show that the magnetic behavior of spin-$(S,s)$ quantum
ferrimagnets is regarded as the combination of those of spin-$(S-s)$
ferromagnets and spin-$(2s)$ antiferromagnets provided $S=2s$.
The dual aspect is qualitatively generic to all the Heisenberg
ferrimagnets, but those with $S=2s$ should quantitatively be
distinguished from the others.

   We investigate the zero-field magnetic susceptibility $\chi$ of
the model (\ref{E:H}) of $N$ elementary cells with $J_1=J_2=J$ under
the periodic boundary condition.
We take $\mu_B$ as the Bohr magneton and set the $g$ factors of both
spins $S$ and $s$ equal to $g$.
We assume that $S>s$.
The numerical procedure of the quantum Monte Carlo method employed
has been detailed elsewhere \cite{Yama64}.
Since the correlation length of the system is considerably small
\cite{Pati94,Breh21}, the thermal quantities in general show no
significant size dependence.
We have calculated the chains of $N=24$ and $N=32$, but we find
no difference beyond the numerical uncertainty between them in the
temperature region treated.
On the other hand, for the sake of argument, we calculate the
susceptibilities of Heisenberg ferromagnetic and antiferromagnetic
rings as well, where we consider chains of $N$ spins.

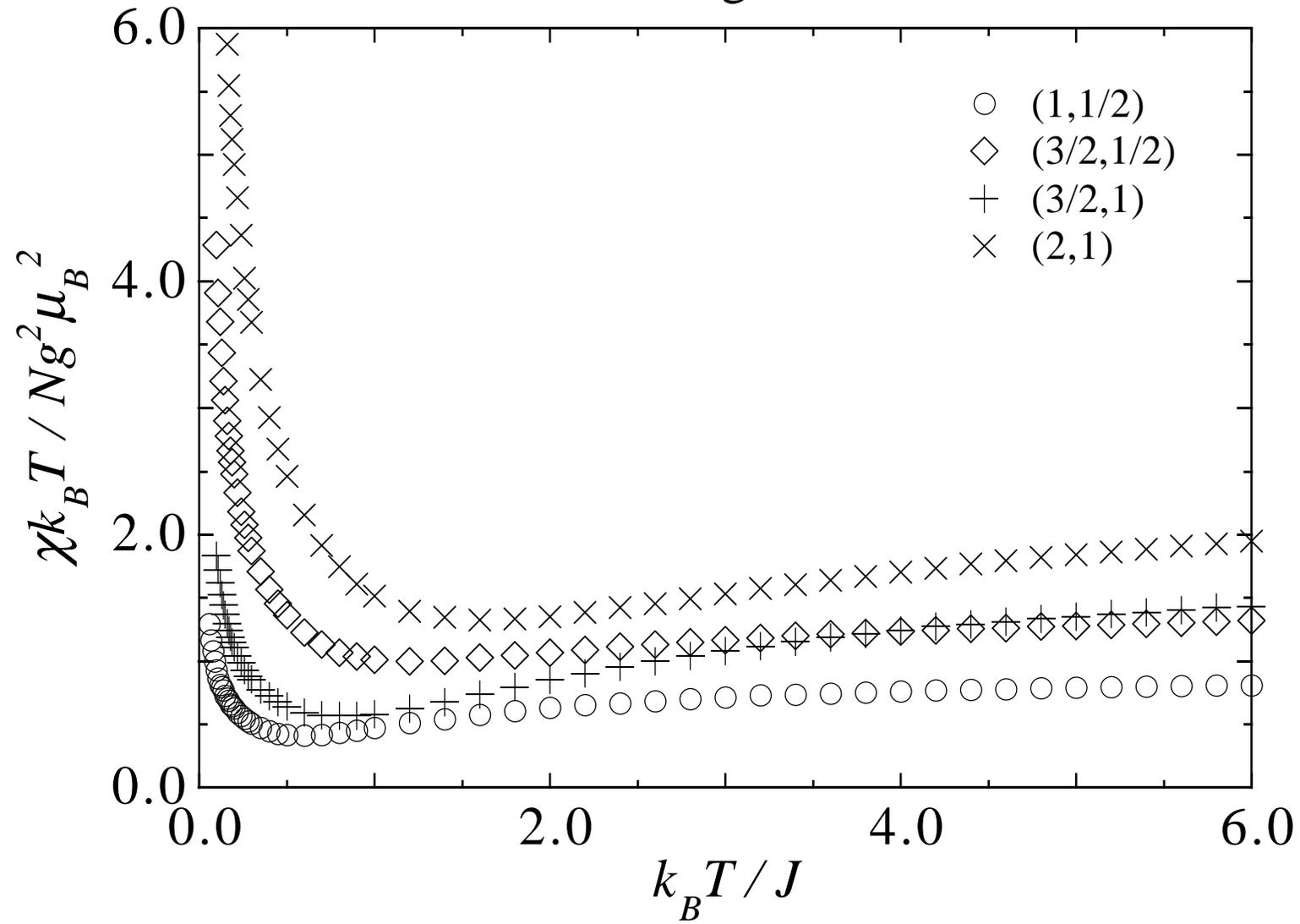
\begin{figure}
\caption{Temperature dependences of the magnetic susceptibility times
         temperature for the Heiseberg ferrimagnetic chains of
         $N=32$.}
\label{F:chiTall}
\end{figure}

   We show in Fig. \ref{F:chiTall} temperature dependences of the
product $\chi T$ at various combinations of $(S,s)$.
We note that here $\chi T$ shows a minimum in its temperature
dependence.
Although the low-temperature divergence, which is proportional to
$T^{-1}$, is reminiscent of the ferromagnetic susceptibility, $\chi T$
approaches its paramagnetic behavior $[S(S+1)+s(s+1)]/3$ showing
the antiferromagnetic increase.
Considering that $\chi T$ is a monotonically decreasing function in
ferromagnets, while a monotonically increasing function in
antiferromagnets, the present observations may be regarded as
a ferromagnetic-antiferromagnetic crossover.

 This mixed feature is a consequence of the dual structure of
excitations \cite{Yama09}.
Quantum ferrimagnets exhibit two distinct branches of elementary
excitations, one of which reduces the magnetization
$M\equiv\sum_j(S_j^z+s_j^z)$ and is thus called ferromagnetic, and the
other of which enhances that and is thus called antiferromagnetic.
The lowest-order spin-wave approach \cite{Pati94,Breh21} to the model
(\ref{E:H}) with $J_1=J_2=J$ indeed results in two dispersions
\begin{equation}
   \omega_k^\mp
     =J\left[
        \sqrt{(S-s)^2+4Ss\sin^2(ak)}\mp(S-s)
       \right]\,,
   \label{E:SWdspmp}
\end{equation}
with the lattice spacing $a$, which are quadratic at small $k$'s and
gapped from the ground state, respectively.
Another naive but suggestive calculation \cite{Yama09} is the
perturbation from the decoupled-dimer limit, where the ground state
and its elementary excitations are trivially obtained
(Fig. \ref{F:illust1-1/2}).
When we turn on the exchange interaction between the dimers, the
localized excitations begin to hop to neighboring unit cells and
become dispersive.
In the case of $(S,s)=(1,1/2)$, for instance, setting $J_1$ and $J_2$
for $J$ and $\delta J$, respectively, we obtain
\begin{eqnarray}
   \omega_k^-
      &=&\frac{4}{9}J\delta[1-\cos(2ak)]
        +O(\delta^2)\,,
   \label{E:pertdspm} \\
   \omega_k^+
      &=&\frac{3}{2}J+\frac{1}{18}J\delta[7-12\cos(2ak)]
        +O(\delta^2)\,.
   \label{E:pertdspp}
\end{eqnarray}
For an arbitrary combination of $(S,s)$, the similar argument ends up
with qualitatively the same dispersions.

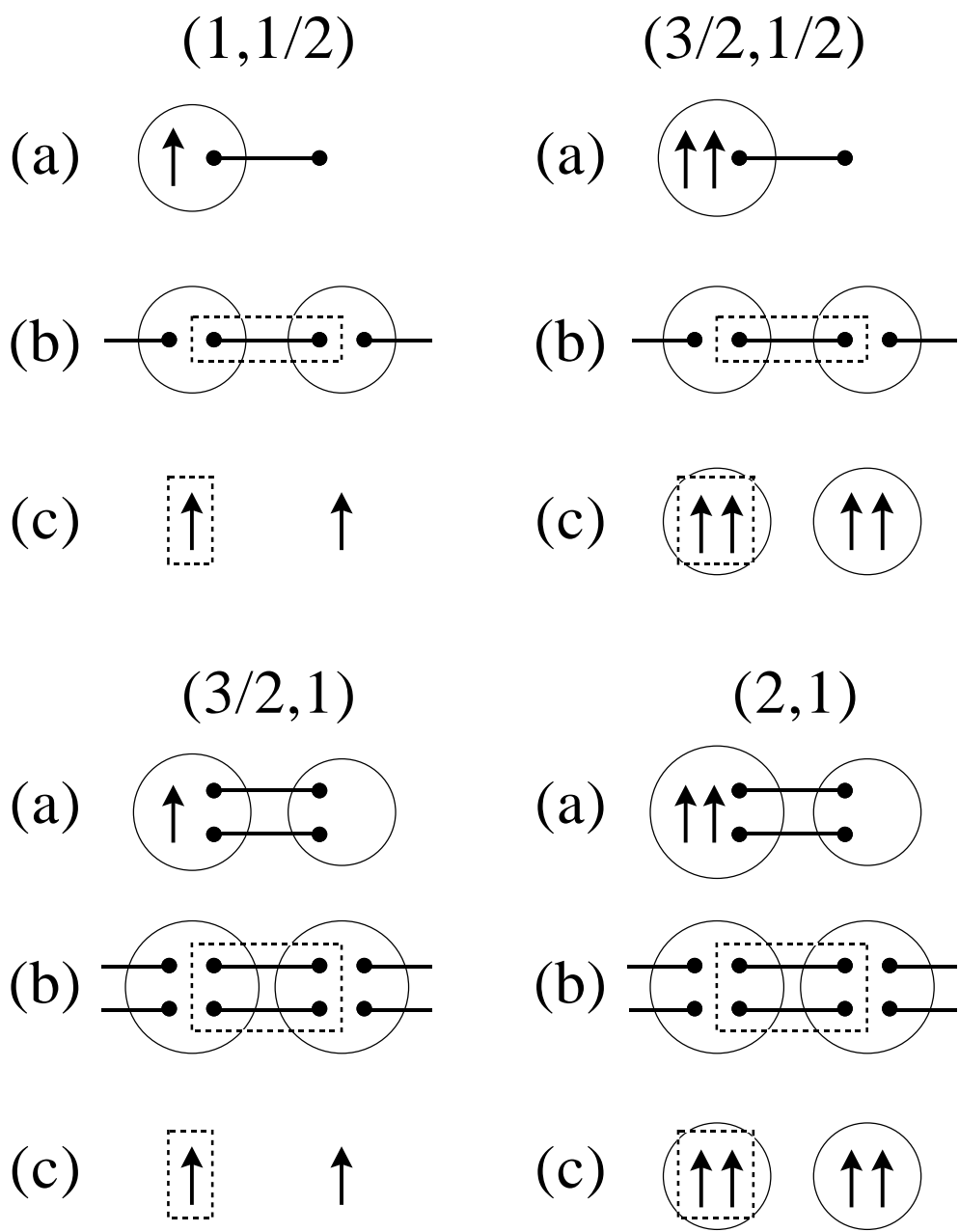
\begin{figure}
\caption{Schematic representations of the $M=(S-s)N$ ground states of
         spin-$(S,s)$ ferrimagnetic chains of $N$ elementary cells in
         the decoupled-dimer limit (a), the AKLT ground states of
         spin-$(2s)$ antiferromagnetic chains of $2N$ spins (b), the
         $M=2N(S-s)$ ground states of spin-$(S-s)$ ferromagnetic
         chains of $2N$ spins (c).
         The notation is the same as in Fig. 2.
         The relation ${\rm(a)}\approx[{\rm(b)}+{\rm(c)}]/2$ may be
         expected.}
\label{F:illustall}
\end{figure}

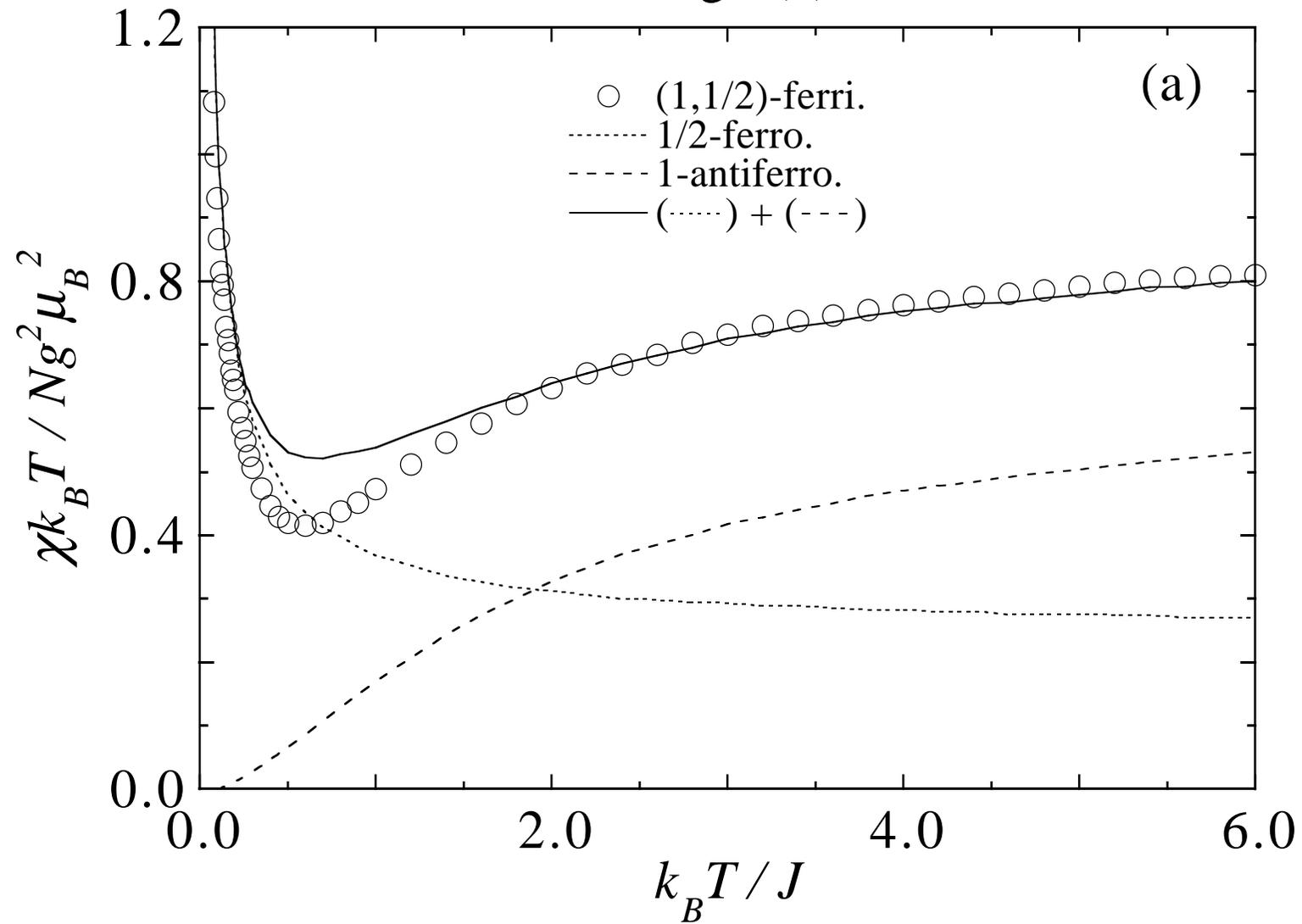
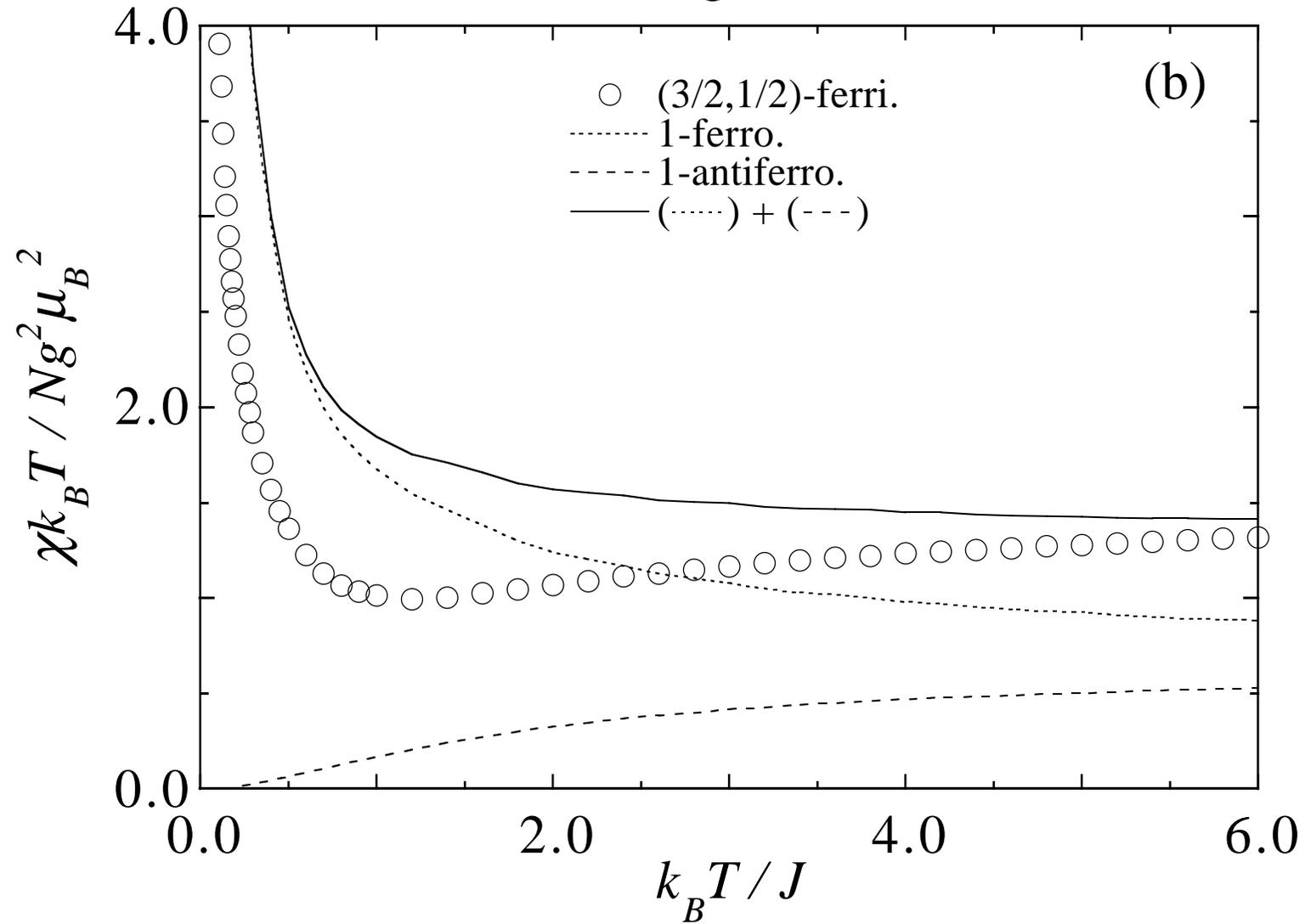
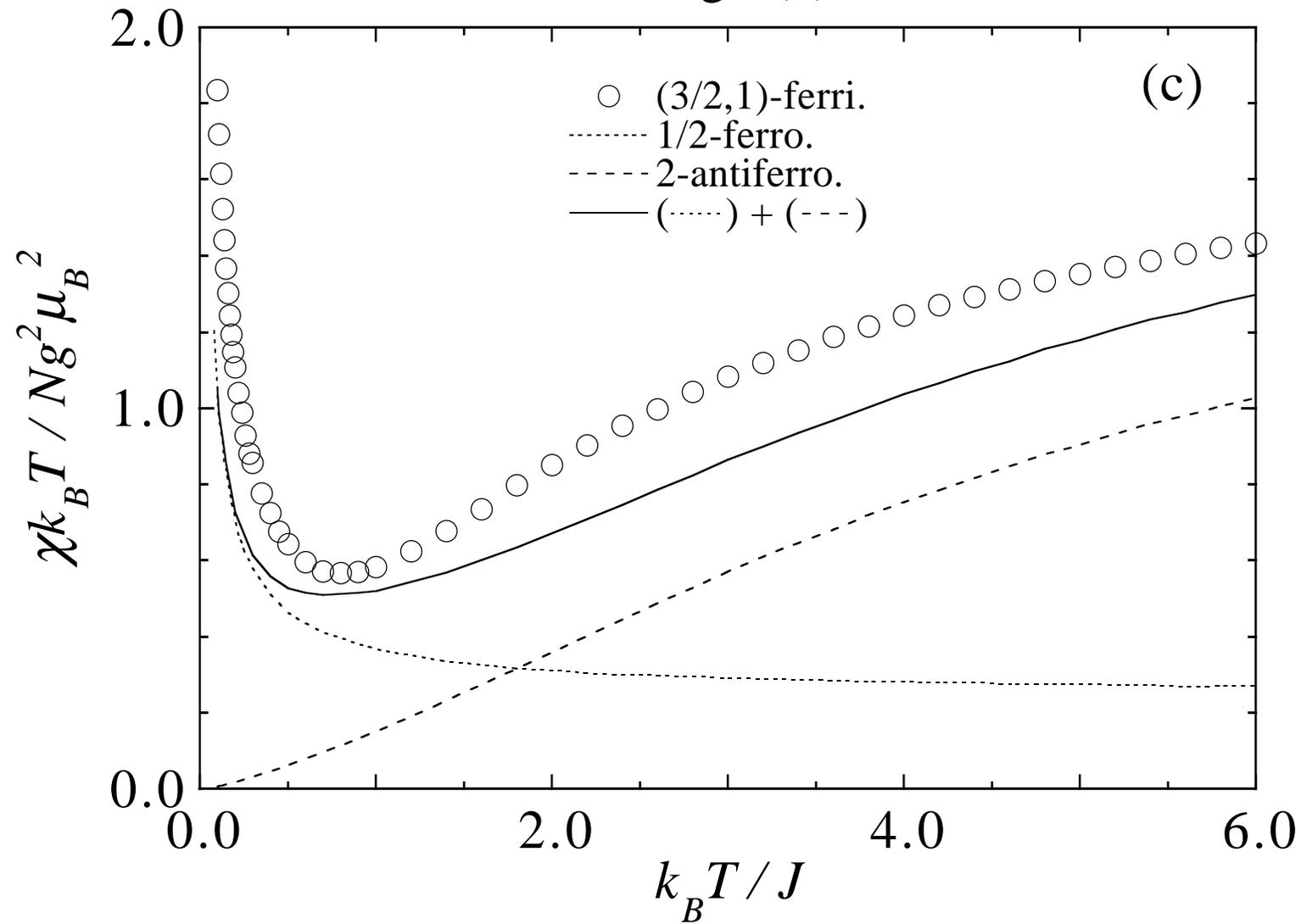
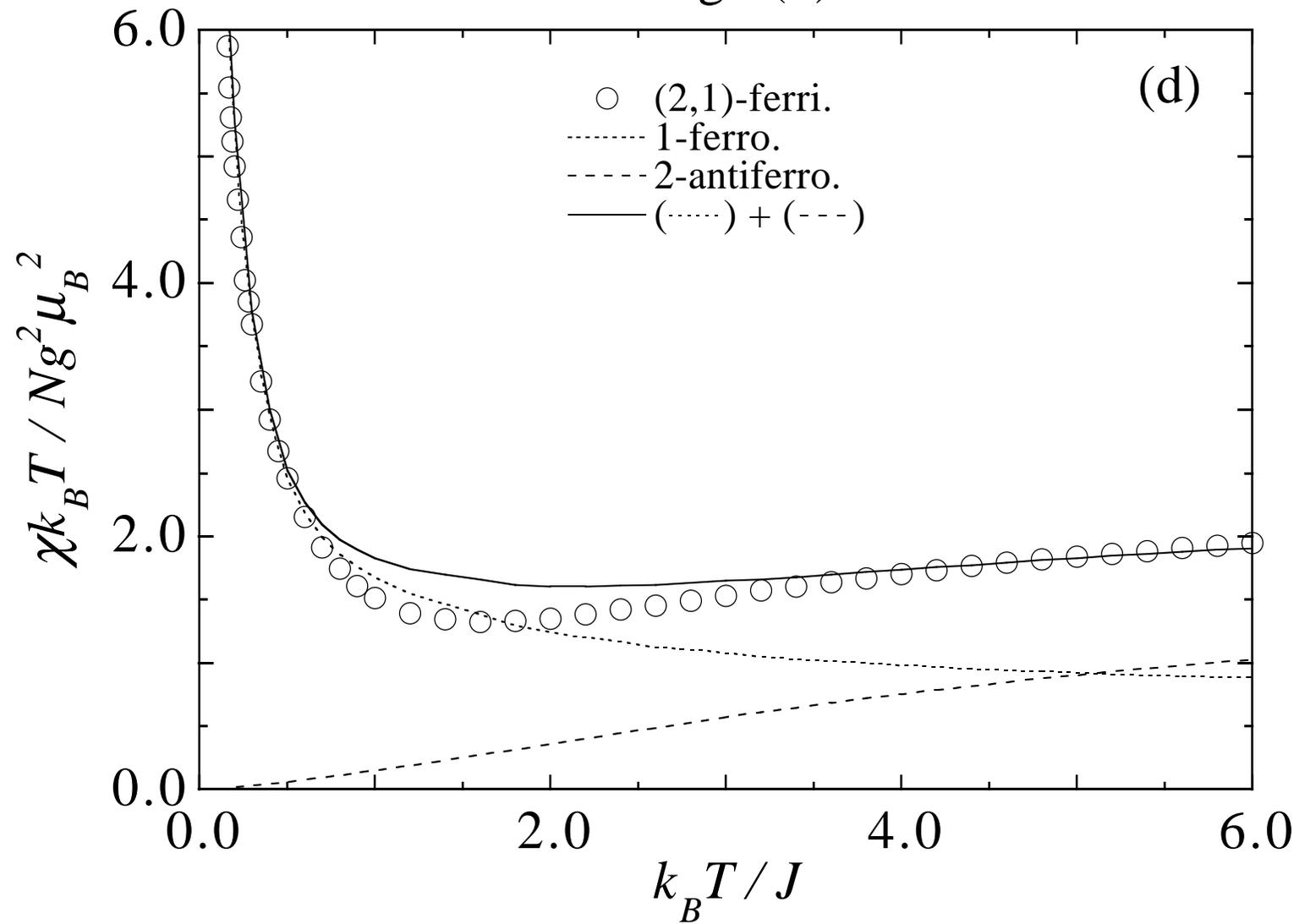
\begin{figure}
\caption{Susceptibilities shown in Fig. 1 are compared with those of
         the spin-$(S-s)$ ferromagnetic and spin-$(2s)$
         antiferromagnetic Heisenberg chains of $N$ spins.}
\label{F:chiTeach}
\end{figure}

   Based on this scenario for low-lying excitations, we propose in
Fig. \ref{F:illustall} an idea of decomposing ferrimagnets into
ferromagnets and antiferromagnets, where we let the decoupled dimers
and the Affleck-Kennedy-Lieb-Tasaki valence-bond-solid states
\cite{Affl99} symbolize ferrimagnets and integer-spin gapped
antiferromagnets, respectively.
Now we are led to expect spin-$(S,s)$ ferrimagnets to behave like
combinations of spin-$(2s)$ antiferromagnets and spin-$(S-s)$
ferromagnets.
In this context, we compare in Fig. \ref{F:chiTeach} the
ferrimagnetic susceptibilities with those of ferromagnets and
antiferromagnets.
Surprisingly, the sum of the spin-$(2s)$ antiferromagnetic and
spin-$(S-s)$ ferromagnetic susceptibilities excellently reproduces the
spin-$(S,s)$ ferrimagnetic behavior on condition that $S=2s$, whereas
such a naive relation does not stand up in the other cases.
Considering several suggestive calculations, we may be convinced of
the present observations.

   First, the high-temperature paramagnetic behavior distinguishes
the $S=2s$ cases from the others.
At high enough temperatures, we should obtain the paramagnetic
susceptibilities as
\begin{eqnarray}
   &&
   \frac{k_{\rm B}T\chi^{(S,s){\scriptstyle\mbox{-}}{\rm ferri.}}}
        {Ng^2\mu_{\rm B}^2}
   =
   \frac{1}{3}
   \left[
    S(S+1)+s(s+1)
   \right]\,,\\
   &&
   \frac{k_{\rm B}T\chi^{(S-s){\scriptstyle\mbox{-}}{\rm ferro.}}}
        {Ng^2\mu_{\rm B}^2}
   =
   \frac{1}{3}(S-s)(S-s+1)\,,\\
   &&
   \frac{k_{\rm B}T\chi^{(2s){\scriptstyle\mbox{-}}{\rm antiferro.}}}
        {Ng^2\mu_{\rm B}^2}
   =
   \frac{1}{3}2s(2s+1)\,.
\end{eqnarray}
It is interesting that we have an equality
$\chi^{(S,s){\scriptstyle\mbox{-}}{\rm ferri.}}
=\chi^{(S-s){\scriptstyle\mbox{-}}{\rm ferro.}}
+\chi^{(2s){\scriptstyle\mbox{-}}{\rm antiferro.}}$
just in the cases of $S=2s$.

   Second, the spin-wave dispersions (\ref{E:SWdspmp}) give us a
useful piece of information.
The small-momentum ferromagnetic excitations must dominate the
low-temperature thermodynamics and therefore the curvature of their
dispersion is of great interest.
Let us recall that spin-$S$ ferromagnetic Heisenberg chains exhibit
the dispersion of single-magnon excitations,
\begin{equation}
   \omega_k=2JS(1-\cos ka)\,.
\end{equation}
Defining the curvature $v$ for quadratic dispersions as
$\omega_{k\rightarrow 0}=v(\widetilde{a}k)^2$
with the unit-cell length $\widetilde{a}$, we obtain
\begin{eqnarray}
   &&
   v^{(S,s){\scriptstyle\mbox{-}}{\rm ferri.}}
     =\frac{Ss}{2(S-s)}J\,,
   \label{E:vferri}\\
   &&
   v^{(S-s){\scriptstyle\mbox{-}}{\rm ferro.}}
     =(S-s)J\,.
   \label{E:vferro}
\end{eqnarray}
Here we learn that
$v^{(S,s){\scriptstyle\mbox{-}}{\rm ferri.}}
=v^{(S-s){\scriptstyle\mbox{-}}{\rm ferro.}}$
provided $S=2s$.
Though the coincidence holds within the lowest-order spin-wave theory,
it looks meaningful.

   Finally, we point out that the modified spin-wave treatment
\cite{Taka33} of spin chains is also suggestive in our argument.
The modified spin-wave theory was first introduced by Takahashi
\cite{Taka68} for quantum ferromagnets.
Imposing a constraint on the magnetization,
he obtained the low-temperature susceptibilities of spin-$S$
ferromagnetic chains as
\widetext
\begin{equation}
   \frac{J\chi^{S{\scriptstyle\mbox{-}}{\rm ferro.}}}
        {N(g\mu_{\rm B})^2}
     =\frac{2S^4}{3}t^{-2}
     -\frac{\zeta(\frac{1}{2})}{\sqrt{\pi}}S^{5/2}t^{-3/2}
     +\left[
       \frac{\zeta(\frac{1}{2})}{\sqrt{\pi}}
      \right]^2
      \frac{S}{2}t^{-1}
      +O(t^{-1/2})
   \,,\label{E:MSWferro}
\end{equation}
where $t=k_{\rm B}T/J$ and $\zeta(z)$ is Riemann's zeta function.
Equation (\ref{E:MSWferro}) was shown to precisely agree with the
thermodynamic Bethe-ansatz calculation \cite{Taka08} in the case of
$S=1/2$.
On the other hand, the author and Fukui \cite{Yama08} have recently
developed the scheme for quantum ferrimagnets.
They constructed the thermodynamics based on the dispersions
(\ref{E:SWdspmp}) and obtained the expression for the susceptibility,
\begin{equation}
   \frac{J\chi^{(S,s){\scriptstyle\mbox{-}}{\rm ferri.}}}
        {N(g\mu_{\rm B})^2}
     = \frac{Ss(S-s)^2}{3}t^{-2}
     -(Ss)^{1/2}(S-s)^{3/2}
      \frac{\zeta(\frac{1}{2})}{\sqrt{2\pi}}t^{-3/2}
     +(S-s)
      \left[
       \frac{\zeta(\frac{1}{2})}{\sqrt{2\pi}}
      \right]^2
      t^{-1}
     +O(t^{-1/2})\,.
   \label{E:MSWferri}
\end{equation}
\narrowtext
%\begin{eqnarray}
%   \frac{J\chi^{(S,s){\scriptstyle\mbox{-}}{\rm ferri.}}}
%        {N(g\mu_{\rm B})^2}
%    &=&\frac{Ss(S-s)^2}{3}t^{-2}
%     -(Ss)^{1/2}(S-s)^{3/2}
%      \frac{\zeta(\frac{1}{2})}{\sqrt{2\pi}}t^{-3/2}
%    \nonumber \\
%    &+&(S-s)
%      \left[
%       \frac{\zeta(\frac{1}{2})}{\sqrt{2\pi}}
%      \right]^2
%      t^{-1}
%     +O(t^{-1/2})\,.
%   \label{E:MSWferri}
%\end{eqnarray}
For ferrimagnets the constraint is not so trivial as for ferromagnets.
However, as far as the low-temperature behavior is concerned, we
obtain the universal description regardless of the artificial
constraint.
Equation (\ref{E:MSWferri}) is thus reliable to a certain extent.
Now we again encounter the surprising coincidence between the two
results,
$\chi^{(S,s){\scriptstyle\mbox{-}}{\rm ferri.}}=
 \chi^{(S-s){\scriptstyle\mbox{-}}{\rm ferro.}}$,
provided $S=2s$.

   Thus we have learned that several analytic treatments, as well as
numerical calculations, distinguish ferrimagnetic chains with $S=2s$
from the others.
We conclude an approximate {\it sum rule} among ferrimagnets,
ferromagnets, and antiferromagnets on condition that $S=2s$:
\begin{equation}
  \chi^{(S,s){\scriptstyle\mbox{-}}{\rm ferri.}}
 =\chi^{(S-s){\scriptstyle\mbox{-}}{\rm ferro.}}
 +\chi^{(2s){\scriptstyle\mbox{-}}{\rm antiferro.}}\,.
  \label{E:Srule}
\end{equation}
Equation (\ref{E:Srule}) well holds in a wide region of temperature
except for the ferromagnetic-to-antiferromagnetic transitional period.
Both ferromagnetic and antiferromagnetic features generically lie in
quantum ferrimagnets, to be sure, but the quantitative identity of
them is established only in the cases of $S=2s$.
For $S>2s$ quantum ferrimagnets are rather ferromagnetic than
antiferromagnetic, whereas for $S<2s$ vice versa.
In fact we find in Fig. \ref{F:chiTall} that $\chi T$ shows a
relatively significant low-temperature divergence for $S>2s$, while it
increases relatively rapidly at intermediate temperatures for $S<2s$.
Brand-new experiments \cite{Hagi09,Fuji} are in progress under the
current interest.
We hope that the second wave of close collaboration between
theoretical and experimental investigations will be brought about.

   It is a pleasure to thank T. Fukui, H.-J. Mikeska, and S. K. Pati
for fruitful discussions.
This work was supported by the Japanese Ministry of Education,
Science, and Culture through Grant-in-Aid No. 09740286 and by the
Okayama Foundation for Science and Technology.
The numerical computation was done in part using the facility of the
Supercomputer Center, Institute for Solid State Physics, University of
Tokyo.

\newpage

\begin{figure}
\caption{Schematic representations of
         the $M=(S-s)N$ ground state of the decoupled dimers
         composed of spins $1$ and $1/2$ and its ferromagnetic (b)
         and antiferromagnetic (c) excitations.
         The arrow (the bullet symbol) denotes a spin $1/2$ with its
         fixed (unfixed) projection value.
         The solid (broken) segment is a singlet (triplet) pair.
         The circle represents an operation of constructing a spin $1$
         by symmetrizing the two spin $1/2$'s inside.}
\label{F:illust1-1/2}
\end{figure}

\widetext

\begin{references}

\bibitem{Dril13}
   M. Drillon, J. C. Gianduzzo, and R. Georges,
      Phys. Lett. {\bf 96A}, 413 (1983);
   M. Drillon, E. Coronado, R. Georges, J. C. Gianduzzo, and
   J. Curely,
      Phys. Rev. B {\bf 40}, 10992 (1989).

\bibitem{Hald64}
   F. D. M. Haldane,
      Phys. Lett. {\bf 93A}, 464 (1983);
      Phys. Rev. Lett. {\bf 50}, 1153 (1983).

\bibitem{Fuku09}
   T. Fukui and N. Kawakami,
      Phys. Rev. B {\bf 55}, 14709 (1997);
      {\it ibid.} {\bf 56}, 8799 (1997).

\bibitem{Taka}
   K. Takano,
      preprint (cond-mat/9804055).

\bibitem{Fuku98}
   T. Fukui and N. Kawakami,
      Phys. Rev. B {\bf 57}, 398 (1998);
   A. Koga, S. Kumada, N. Kawakami, and T. Fukui,
      J. Phys. Soc. Jpn. {\bf 67}, 622 (1998).

\bibitem{Glei73}
   A. Gleizes and M. Verdaguer,
      J. Am. Chem. Soc. {\bf 103}, 7373 (1981):
      {\it ibid.} {\bf 106}, 3727 (1984).

\bibitem{Pei38}
   Y. Pei, M. Verdaguer, O. Kahn, J. Sletten, and J.-P. Renard,
      Inorg. Chem. {\bf 26}, 138 (1987);
   O. Kahn, Y. Pei, M. Verdaguer, J.-P. Renard, and J. Sletten,
      J. Am. Chem. Soc. {\bf 110}, 782 (1988);
   P. J. van Koningsbruggen, O. Kahn, K. Nakatani, Y. Pei, and
   J.-P. Renard,
      Inorg. Chem. {\bf 29}, 3325 (1990).

\bibitem{Dril53}
   M. Drillon, E. Coronado, D. Beltran, R. Georges,
      J. Appl. Phys. {\bf 57}, 3353 (1985);
   M. Drillon, E. Coronado, D. Beltran, J. Curely, R. Georges,
   P. R. Nugteren, L. J. de Jongh, and J. L. Genicon,
      J. Magn. Magn. Mater. {\bf 54-57}, 1507 (1986).

\bibitem{Alca67}
   F. C. Alcaraz and A. L. Malvezzi,
      J. Phys. A {\bf 30}, 767 (1997).

\bibitem{Pati94}
   S. K. Pati, S. Ramasesha, and D. Sen,
      Phys. Rev. B {\bf 55}, 8894 (1997);
      J. Phys.: Condens. Matter {\bf 9}, 8707 (1997).

\bibitem{Kole36}
   A. K. Kolezhuk, H.-J. Mikeska, and S. Yamamoto,
      Phys. Rev. B {\bf 55}, 3336 (1997).

\bibitem{Breh21}
   S. Brehmer, H.-J. Mikeska, and S. Yamamoto,
      J. Phys.: Condens. Matter {\bf 9}, 3921 (1997).

\bibitem{Nigg31}
   H. Niggemann, G. Uimin, and J. Zittartz,
      J. Phys.: Condens. Matter {\bf 9}, 9031 (1997);
      preprint (cond-mat/9712202).

\bibitem{Yama09}
   S. Yamamoto,
      Int. J. Mod. Phys. C {\bf 8}, 609 (1997);
   S. Yamamoto, S. Brehmer, and H.-J. Mikeska,
      Phys. Rev. B {\bf 57}, 13610 (1998).

\bibitem{Ono76}
   T. Ono, T. Nishimura, M. Katsumura, T. Morita, and M. Sugimoto,
      J. Phys. Soc. Jpn. {\bf 66}, 2576 (1997).

\bibitem{Kura29}
   T. Kuramoto,
      J. Phys. Soc. Jpn. {\bf 67}, 1762 (1998).

\bibitem{Yama08}
   S. Yamamoto and T. Fukui,
      Phys. Rev. B {\bf 57}, 14008 (1998).

\bibitem{Ivan14}
   N. B. Ivanov,
      Phys. Rev. B {\bf 57}, June 1 (1998).

\bibitem{Mais}
   K. Maisinger, U. Schollw\"ock, S. Brehmer, H.-J. Mikeska, and
   S. Yamamoto,
      Phys. Rev. B {\bf 58}, No. 10 (1998).

\bibitem{Affl99}
   I. Affleck, T. Kennedy, E. H. Lieb, and H. Tasaki,
      Phys. Rev. Lett. {\bf 59}, 799 (1987);
      Commun. Math. Phys. {\bf 115}, 477 (1988).

\bibitem{Yama64}
   S. Yamamoto,
      Phys. Rev. B {\bf 53}, 3364 (1996).

\bibitem{Taka33}
   M. Takahashi,
      Prog. Theor. Phys. Suppl. {\bf 87}, 233 (1986).

\bibitem{Taka68}
   M. Takahashi,
      Phys. Rev. Lett. {\bf 58}, 168 (1987).

\bibitem{Taka08}
   M. Takahashi and M. Yamada,
      J. Phys. Soc. Jpn. {\bf 54}, 2808 (1985);
   M. Yamada and M. Takahashi,
      {\it ibid.} {\bf 55}, 2024 (1986).

\bibitem{Hagi09}
   M. Hagiwara, K. Minami, Y. Narumi, K. Tatani, and K. Kindo,
      J. Phys. Soc. Jpn. {\bf 67}, 2209 (1998).

\bibitem{Fuji}
   N. Fujiwara and M. Hagiwara,
      private communication.

\end{references}
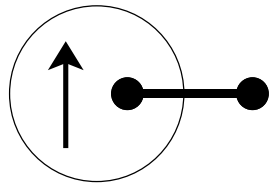
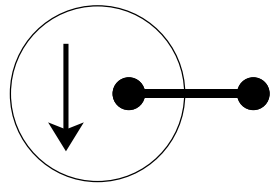
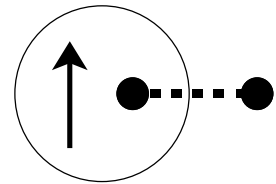
\end{document}